\documentclass[12pt,preprint]{aastex}

\usepackage{natbib}
\usepackage{epsfig}
\def\rhop{\rho_{\rm p}}
\def\rhog{\rho_{\rm g}}
\def\cg{c_{\rm g}}
\def\rhos{\rho_{\rm s}}
\def\Sigp{\Sigma_{\rm p}}
\newcommand{\Sigpc}{\Sigma_{\mathrm{p, cr}}}
\def\Sigg{\Sigma_{\rm g}}
\def\sp{_{\rm p}}
\def\g{_{\rm g}}
\def\p{\partial}
\def\l{\left}
\def\r{\right}
\def\ts{t_{\rm stop}}

\def\t{_{\mathrm{turb}}}
\newcommand{\vt}{v_{\mathrm{r,turb}}}

\newcommand{\HEk}{H_{\mathrm{Ek}}} 

\newcommand{\gcc}{\;\mathrm{g/cm^{3}}}

\newcommand{\cm}{\; {\rm cm}}
\newcommand{\mm}{\; {\rm mm}}

\newcommand{\AU}{\; {\rm AU}}

\newcommand{\be}{\begin{equation}}
\newcommand{\ee}{\end{equation}}
\newcommand{\bea}{\begin{eqnarray}}
\newcommand{\eea}{\end{eqnarray}}

\begin{document}
\title{Particle Pile-ups and Planetesimal Formation}

\author{Andrew N. Youdin
\& Eugene I. Chiang}

\affil{
Department of Astronomy, University of California, Berkeley, CA 94720}

\begin{abstract}
Solid particles in protoplanetary disks that are
sufficiently super-solar in metallicity overcome turbulence generated
by vertical shear to gravitationally
condense into planetesimals.
Super-solar metallicities result if solid particles pile
up as they migrate starward as a result of
aerodynamic drag. Previous analyses of aerodynamic drift rates
that account for mean flow differences between gas and particles
yield particle pile-ups. We improve on
these studies not only by accounting for the collective
inertia of solids relative to that of gas, but also by
including the transport of angular momentum by
turbulent stresses within the particle layer.
These turbulent stresses are derived in a physically
self-consistent manner from the structure of
marginally Kelvin-Helmholtz turbulent flows.
They are not calculated using the usual plate drag formulae,
whose use we explain is inappropriate.
Accounting for the relative inertia of solids to gas
retards, but does not prevent, particle pile-ups,
and generates more spatially extended regions of
metal enrichment. Turbulent transport hastens
pile-ups. We conclude
that particle pile-up is a robust outcome in
sufficiently passive protoplanetary disks.
Connections to observations of circumstellar
disks, including the Kuiper Belt, and the architectures of planetary systems
are made.
\end{abstract}


\section{Introduction}
\label{introd}
The formation of planetesimals by gravitational instability within
dense particle layers in circumstellar disks
is an appealing alternative to collisional sticking of dust aggregates
(\citealp{ys02}, hereafter YS02, and references therein).
In one fell swoop, gravity promises to assemble sand-sized
or smaller solids into kilometer-sized agglomerates.

For gravitational instability to operate, two main requirements
must be met. First, the disk must be sufficiently passive;
turbulent velocity fluctuations in disk gas must be small enough
to allow solids to gravitationally settle toward the midplane.
Kelvin-Helmholtz turbulence generated by vertical
shear within the stratified particle
layer threatens to violate this requirement
\citep{cdc93, stu95}.  Such turbulence, however, can entrain only
a finite amount of solids.
Thus, a second requirement for gravitational instability,
first explored by \citet{sek98}, is that the surface
density of solids relative to that of gas, $\Sigp/\Sigg$,
must exceed a critical threshold, $\Sigpc/\Sigg$.
Within the minimum-mass solar nebula (MMSN),
$\Sigpc/\Sigg$ lies $\sim$5 --- 20 times
above the solar value of $\Sigp/\Sigg \sim 5 \times 10^{-3}$.

Many possibilities exist for achieving this enhancement
of metallicity, including the
return of condensed material to an accretion disk from a bipolar outflow
and removal of dust-depleted gas in disk surface layers
by magnetic accretion, photoevaporation, or entrainment
within a stellar wind.
YS02 discuss the origins and merits of these possibilities.
They highlight a promising and natural enrichment mechanism that arises
from simple aerodynamic drag. Solid particles rotate about the central star
faster than does ambient gas because the latter is more
sensitive to pressure gradients that are
(usually) directed radially outward.
This difference in mean flow velocities causes
particles to be frictionally dragged by gas. Such
drag induces orbital decay.
By applying the Epstein aerodynamic drag law to millimeter-sized
solids in the
MMSN, YS02 conclude
that as particles drift relative to gas toward the central star,
they pile up: local enhancements
of more than an order-of-magnitude in the particle surface density
(metallicity) occur at small stellocentric radii.

The Epstein drag law applies when
relative velocities, $v_{\rm{rel}}$, between a solid particle
and surrounding gas
are less than the gas sound speed, $\cg$,
and when
the particle radius, $a$, is smaller than
the gaseous mean free path, $\lambda$. These conditions
are satisfied for particles having $a \lesssim 1\,(r/{\rm AU})^{2.75}\cm$
in the MMSN, where $r$ is the stellocentric distance.
The Epstein
drag force on a spherical particle reads

\begin{equation}
D_{\rm Ep} = {4 \pi \over 3} \rhog \cg v_{\rm{rel}} a^2 \, ,
\label{beginepstein}
\end{equation}
\noindent where $\rhog$ is the mass density of gas.
YS02 find that millimeter-sized particles pile up
to the point where their density exceeds the Roche
density in $\sim$$10^5$ yr, fast enough to
occur within disk lifetimes ($\sim$$10^7$ yr).

Recently, \citet{stu03} has argued that metallicity enhancement
is unlikely because Epstein-type drag,
as described above, inadequately describes
the forces on particles.
Since the timescale for a particle to settle vertically toward
the midplane is
shorter than the timescale for a particle to drift radially
by Epstein drag [by a factor of order
$\eta \sim 10^{-3} (r/\rm{AU})^{1/2}$; see the next section],
particles settle vertically
into states of marginal Kelvin-Helmholtz turbulence (see Sekiya 1998)
before drifting appreciably in the radial direction.
Such turbulence exerts additional stresses
on particles, stresses that are not accounted for by Epstein drag;
radial drift rates could, in principle, be altered significantly.
Weidenschilling (2003) proceeds by assuming, as have other authors in different
contexts (e.g., \citealp{gw73,gp00}, hereafter GP00), that this
extra turbulent stress
can be modeled as ``plate drag'': the turbulent stress on the
particle layer is taken to be akin to the turbulent stress on a solid,
rigid plate that is dragged through gas. The turbulent
stresses so prescribed still induce inward radial drift of solids
relative to gas, but the new, equilibrium
drift velocities scale with stellocentric
distance in such a way that particle pile-ups and
metallicity enhancements do not occur in the MMSN.

GP00 also employ the plate drag prescription to study the
local, two dimensional dynamics of the turbulent dust sheet.
They discover an instability that causes in-plane perturbations to grow
over the local dynamical time.
This instability was not recovered by Weidenschilling (2003) because
the latter solved for the equilibrium state, in which
unbalanced gravitational forces on the dust layer
match exactly the drag force;
perturbations in radial momentum about this state were
not considered.\footnote{These
equilibrium states are referred to as
``constant states'' by GP00.}

Our purpose here is to discard the usual plate drag prescription
in favor of a self-consistent calculation of turbulent stresses
based on the structure of the stratified particle layer.
The essence of our approach is to solve for the turbulent
diffusivity necessary to keep solids mixed vertically
in the Kelvin-Helmholtz turbulent state, and then to employ
this diffusivity in calculating rates of angular momentum
transport.
We find that the inclusion of turbulent stresses
promotes local metal enrichment as particles
accrete toward the central star; particles pile up radially
more quickly with such stresses than without them.

Our analysis below is restricted to millimeter-sized particles,
though it can be extended to govern smaller sizes. As long as
the Schmidt numbers of individual particles are
nearly unity (see \S\ref{newcalc}), the particles are well entrained
in the gas and we may model the gas and particles as a single fluid.
One millimeter is an interesting size regime from a number
of perspectives. First, insofar as we would like to form
planetesimals from seed particles that are as small as possible, with
minimum recourse to collisional sticking, millimeter (or smaller)
scales are more interesting to us than the meter scales
that concern other works such as GP00. Second, collisional
sticking might stall at sizes of a few millimeters,
since relative velocities between particles tend to increase
with increasing particle size, and relative velocities
that are too high result in shattering of particle aggregates
rather than sticking. Current consensus tells us that
the threshold velocity that divides sticking from shattering
is $\sim1~{\rm m~s}^{-1}$ and that
the corresponding maximum particle size generated by
particle-particle collisions is $\sim 1$ cm (Weidenschilling \&
Cuzzi 1993; Blum \& Wurm 2000; Wurm, Blum, \& Colwell 2001; Chiang 2003).
Third, modelling of millimeter-wave spectra of T Tauri and Herbig
Ae star-disk systems offers evidence in favor of millimeter-sized grains
dominating the particle mass near disk midplanes;
the evidence is critically reviewed
by Chiang (2003). Fourth, millimeter sizes characterize
chondrules, the dominant constituent of the most primitive
meteorites.

In \S\ref{mech} we describe aerodynamic drift mechanisms for solids.
Epstein drag is recapitulated in \S\ref{drag},
while turbulent stresses are considered
in \S\ref{TS}. We review the plate drag prescription in
\S\ref{sss:PD} and introduce our new, self-consistent approach
for calculating the transport of angular momentum
by Kelvin-Helmholtz turbulence
in \S\ref{newcalc}. In \S\ref{global} we apply our theory to a
global simulation of the evolution of $\Sigp$ and demonstrate
how particles pile up robustly. In \S\ref{PD} we argue
more pointedly why our treatment of turbulent drag represents
an improvement over the usual plate drag approximation.
Concluding remarks are made in
\S\ref{conc}.  Related discussions of the character
and strength of turbulence in
planetesimal forming disks are relegated to the appendices.

\section{Mechanisms for Radial Drift of Solids}\label{mech}
\subsection{Epstein Drag}\label{drag}

Gas in protoplanetary disks rotates at speeds lower
than the Keplerian speed, $v_K$, by an amount $\sim$$\eta v_K$, where
\be
\eta \equiv -{\p P\g/\p \ln r \over 2 \rho\g v_K^2} \approx \left({c_{\rm g}
\over
v_K}\right)^2
\ee
is a dimensionless measure of radial pressure support
and $P\g$ is the gas pressure.
Typical models of the MMSN are characterized by $\eta \sim 10^{-3}
(r/\rm{AU})^{1/2}$.

The Epstein aerodynamic drag law [equation (\ref{beginepstein})]
implies that the stopping time of a particle of mass $m_{\rm p}$
moving relative to gas is

\be
\label{tsjeremy}
\ts \equiv \frac{m_{\rm p} v_{\rm{rel}}}{D_{\rm Ep}} = \frac{\rhos a}{\rhog
c\g} \, .
\ee
For a particle of internal density $\rhos = 3 \gcc$ and
radius $a = 1$ mm, the stopping time is likely shorter
than the disk rotation period, $r/v_K = 1/\Omega$; to wit,

\be
\Omega \ts \sim 10^{-4} \left( \frac{r}{{\rm AU}} \right)^p \, ,
\ee
where the index $p$
describes the fall-off of gas surface density with distance,

\be\label{sigg}
\Sigg \propto r^{-p} \, .
\ee
If $\Omega \ts \ll 1$, gas and solids rotate at nearly---to order $\eta(\Omega
\ts)^2$---the same azimuthal velocity,
\be\label{vp}
v_\phi = \left[1-\eta {\rhog \over \rhog + \rhop(z)}\right]v_K
\ee
\citep{nak86}.
We make explicit the dependence of particle mass density, $\rhop$,
on vertical height, $z$, above the midplane
because vertical stratification of solids provides the only
significant source of vertical shear, $\p v_\phi / \p z$.
This shear, in turn, drives Kelvin-Helmholtz turbulence.

The radial drift speed of an individual particle
can be derived via force balance in the radial and azimuthal
directions (see, e.g., Goldreich \& Ward 1973). Under the approximation that
the azimuthal velocity of the particle equals the azimuthal
velocity of the gas, the equation for radial force balance
yields for the radial drift speed:

\be\label{ve}
v_{\rm Ep, ind} = 2 \eta \Omega \ts v_K \,.
\ee
The convention in this paper is that radially
inward velocities are positive. Corrections to this
expression are higher order in $\Omega \ts$.
When the inertia of a collection of solids is taken into account,
equation (\ref{ve}) becomes
\be\label{ve'}
v_{\rm{Ep}} = \left({\rhog \over \rho}\right)^2 v_{\rm Ep,ind},
\ee
which is always smaller than $v_{\rm Ep,ind}$ because $\rho = \rhop + \rhog$
\citep{nak86}.
We refer to this dampening of the drift rate due to the collective
inertia of the solids as the  ``inertial slow-down'' effect.
Note that while we have subscripted our velocities in equations
(\ref{ve}) and (\ref{ve'}) by ``Ep'' to denote the Epstein drag law,
the form of the right-hand-side of equation (\ref{ve})
and the correction factor of $\rhog/\rho$ in equation
(\ref{ve'}) are independent of the drag law employed
(e.g., for $a > \lambda$, the appropriate drag law
is due to Stokes: $D \propto a \lambda \cg v_{\rm rel}$).
The same independence does not apply to the explicit form of $\ts$,
which is given for the specific case of Epstein drag on
the far right-hand side of equation (\ref{tsjeremy}).

Particle pile-ups occur if the mass accretion rate in particles
decreases with decreasing $r$ at fixed time. Define

\be
\Sigp \propto r^{-n}
\ee
to be the surface density of solids. Further define the radial drift
rate of solids to be, in general,

\be
v_r \propto r^{d} \, .
\ee
Then the mass accretion rate in particles scales as

\be
\Sigp r v_r \propto r^E \, ,
\ee
where $E = d-n+1$. If $E > 0$, then
accretion results in particle pile-ups and local metal enrichment.

For the specific case of Epstein drag,

\be
\label{depsform}
d_{\rm{Ep}} = \frac{1}{2} - q + p \, ,
\ee
\be
E_{\rm{Ep}} = \frac{3}{2} + p - n - q \, ,
\ee
where the small inertial slow-down correction has been
neglected. Here $q$
describes the fall-off of midplane gas temperature with distance,

\be \label{T}
T \propto r^{-q} \, .
\ee
Reasonable values of $q$ (see, e.g., Chiang \& Goldreich 1997)
span the range of 0.43--0.75. If the disk begins in a well-mixed
state for which $p=n$, then $E_{\rm Ep} \gtrsim 1$ initially and particles pile
up as
they accrete.  Pile-ups cannot occur everywhere for all time.
When a realistic disk with finite outer radius is considered,
$\Sigp$ at a fixed $r$ will grow until particles at the outer radius drift past
$r$.

\begin{figure}[tb]
\centerline{\epsfig{figure=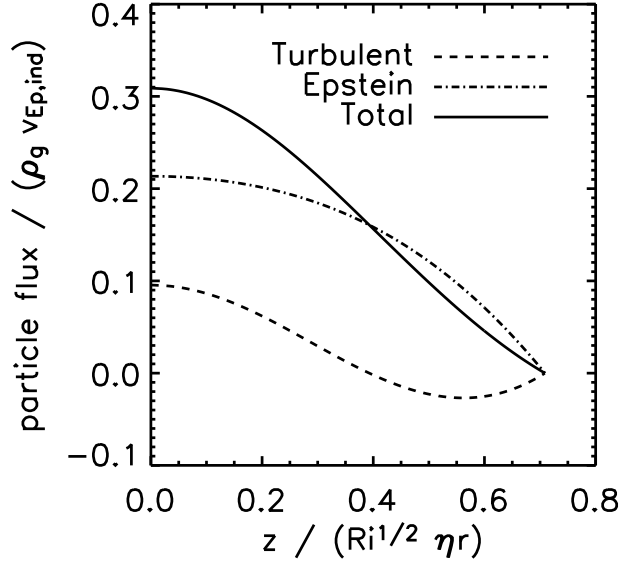, width=9.2cm}}
\caption{Radial particle fluxes from turbulent stresses,
Epstein drag, and their sum vs.\ height above the midplane
for millimeter-sized particles in model H at $r \sim 1 \AU$.
The vertically integrated abundance of chondrule-like particles
is fixed at its cosmic value. Positive fluxes represent inflow.}\label{fluxH1}
\end{figure}

Figure \ref{fluxH1} portrays the particle flux from Epstein drag,
\be\label{fE}
f_{\rm Ep} = \rhop v_{\rm{Ep}} \, ,
\ee
as a function of $z$ for particles having $a = 1 \mm$
and $\rhos = 3 \gcc$ at $r = 1 \AU$ in Hayashi's (1981) model of the
MMSN. We refer to this background nebular
model hereafter as ``model H.''  See YS02
for details of all nebular
models used in this paper. Figure \ref{H1crit}
is analogous to Figure \ref{fluxH1}, except that $\Sigp/\Sigg$ is set
to near the saturation limit, $\Sigp/\Sigg = 0.99 \Sigpc/\Sigg$;
for the case of model H at $r = 1 \AU$, this represents a factor
of $\sim$17 enhancement over solar metallicity.
Distributions of particle density with height are computed
using the model of \citet{sek98} for which the Richardson number equals
1/4 everywhere. The fluxes are scaled
to $\rhog v_{\rm Ep,ind}$, which is independent of $\rhop$.

Before discussing the turbulent contribution to radial drift rates
of solids, we prove that vertical settling of particles
into the Kelvin-Helmholtz turbulent state occurs well before
particles drift radially inward by Epstein drag.
The timescale for vertical settling equals
\be
\label{tsettle}
t_{\rm sett} \sim \frac{H\g}{\Omega^2 z t_{\rm stop}} \sim \frac{1}{\Omega^2
t_{\rm stop}} \, ,
\ee
where $H\g = c\g / \Omega$ equals the vertical
scale height of the gas, $\Omega^2 z$ equals the vertical component
of stellar gravity, and we have set $z = H\g$ in the last equality
in working to order-of-magnitude. The timescale for radial drift equals
\be
\label{tdrift}
t_{\rm Ep} \sim \frac{r}{v_{\rm Ep}} \, .
\ee
Substituting (\ref{ve}) and (\ref{ve'}) into (\ref{tdrift})
reveals immediately that
$t_{\rm sett}/t_{\rm Ep} \sim \eta (\rhog/\rho)^2 \ll 1$.


\subsection{Turbulent Stresses in Stratified Disks}\label{TS}
Forces exerted by gas turbulence are often modeled using the
symmetric pressure tensor, $\mathbf{P}$. In dynamical
equilibrium, the advective transport of angular momentum
balances the torque exerted by turbulent pressure fluctuations; to wit,

\be
\label{L} {\vt }{\p \Omega r^2 \over \p r} = -{r \over
\rho}\left({1 \over r^2}{\p r^2 P_{r\phi} \over \p r} + {1 \over r}{\p
P_{\phi \phi} \over \p \phi} + {\p P_{z \phi} \over \p z}\right) \, ,
\ee
where $\vt$ is the radial accretion velocity induced by turbulent stresses and
$(r,\phi,z)$ are the usual cylindrical coordinates. Again, $\vt > 0$ for
radially inward accretion. The term proportional to $P_{r \phi}$
dominates in active accretion disks that are rendered
turbulent by, e.g., magneto-rotational instability \citep{bh91}
or gravitational instability \citep{gam01}.
The disks considered here derive their turbulence
from vertical shear instabilities, so that we set
$P_{r\phi} = 0$ (see the appendices for quantitative
justification).
We further assume that the
disk is axisymmetric so that $\partial P_{\phi\phi}/\partial \phi = 0$.
Equation (\ref{L}) then reduces to

\be
\label{LL}
{\vt \over r}{\p \Omega r^2 \over \p r} = -{1 \over \rho}{\p P_{z \phi} \over
\p z} \, .
\ee

How can one model $P_{z\phi}$ in turbulent particle disks?
In the next two subsections, we examine
two prescriptions: one based on the conventional ``plate drag''
{\it Ansatz}, and a second rooted in the physics of stratified particle layers.



\subsubsection{The Plate Drag Approximation}\label{sss:PD}
We review the ``plate drag'' treatment of turbulent stresses to connect
to previous works [e.g., Goldreich \& Ward (1973); GP00;
Weidenschilling (2003)] and to contrast it with the new approach
we champion in the next section \S\ref{newcalc}.
Section \ref{PD} criticizes the plate drag approximation more directly.

Users of the plate drag prescription do not resolve the stress, $P_{z\phi}$,
as a function of $z$. Instead, $P_{z\phi}$ is vertically
averaged and utilized to calculate an average accretion velocity.
Averaging equation (\ref{LL}) over $z$ yields
\be \label{plateL}
{\langle \vt \rangle  \over r}{\p \Omega r^2 \over \p r} = {S \over
\Sigma(H\sp)} \, ,
\ee
where $\langle \rangle$ denotes a density-weighted average,
$\Sigma(H\sp)$ is the surface density of solids and gas within the
particle layer of half-thickness $H\sp$, and
\be
S = 2\int_0^{H\sp}{\p P_{z\phi} \over \p z}dz = 2P_{z\phi}(z=H\sp).
\ee
We have used the fact that $P_{z\phi}$ vanishes for $|z| > H\sp$
and is an odd function of $z$. The heart of the
plate drag prescription lies in setting
the drag force per unit area, $S$, equal to

\be\label{S}
S = -\rhog {(\Delta v_\phi)^2 \over {\rm Re}^*} \, ,
\ee
where $\Delta v_{\phi} \approx \eta v_K$ is the characteristic azimuthal
velocity difference
between a dense particle-dominated midplane and overlying particle-free gas.

Prescription (\ref{S}) is useful only to the extent that
${\rm Re}^*$ and any radial variations in this quantity are
known. \citet{gw73} assume
that ${\rm Re}^* = 500$, citing an analysis
of tidal boundary layers on the Earth's ocean floor.
More recently, \citet{cdc93} and \citet{dob99}
obtain values of ${\rm Re}^*$ ranging anywhere from 20 to 180
by using (the square of)
three experimentally determined boundary lengths.
Whether the results of these terrestrial-scale experiments
can be extrapolated to astronomical settings has not been rigorously
demonstrated.

Insertion of (\ref{S}) into (\ref{plateL}) yields the equilibrium
accretion velocity
\be\label{vpl}
\langle \vt \rangle
= v_{\rm plate} = {2 \rho_g \eta^2 \Omega r^2 \over \Sigp {\rm Re}^*} \, ,
\ee
where we have replaced $\Sigma(H\sp)$ with $\Sigp$ for consistency with
other authors (although this replacement is strictly valid only if
$\rhop \gg \rhog$) and for ease when making analytic scalings.
Equation (\ref{vpl}) implies that the indices $d$ and $E$ introduced
in \S\ref{drag} take the values
$d_\mathrm{plate} = 1 - 3q/2 - p + n$
and $E_\mathrm{plate} =  2-3q/2-p$, respectively,
if we assume that ${\rm Re}^*$ is spatially constant.

In model H, $p = n = 3/2$ and $q = 1/2$.
Then $E_\mathrm{plate} = -1/4$; solids drain from the disk onto the
star and do not pile up. This is the result obtained by
\citet{stu03}, who went on to show
that adding the particle flux due to Epstein drag
to that due to plate drag caused metal enhancement and depletion effects
to cancel nearly exactly. We offer two comments regarding
these results. The first is that whether the equilibrium
solutions yield particle pile-ups is highly model-dependent.
For example, an initial surface density
profile for which $p = n = 1$ and $q = 1/2$ gives
$E_{\rm plate} = 1/4$.\footnote{For particles larger
than the gas mean free path, the drag law is due to
Stokes (see \S\ref{introd}), and $E = -q/2 - n$, i.e.,
no pile-up for decreasing temperature and density profiles.
Loss of such large particles to accretion
onto the star is another reason why we restrict ourselves
to the millimeter sizes.}
Our second comment is that such equilibrium solutions
neglect the instability
discovered by GP00, which yields metallicity enhancement
on dynamical timescales
when the plate drag prescription applies.

In the next section we derive an alternative prescription
for turbulent stresses based on the structure of Kelvin-Helmholtz
turbulent flows.

\begin{figure}[bt]
\centerline{
\epsfig{figure=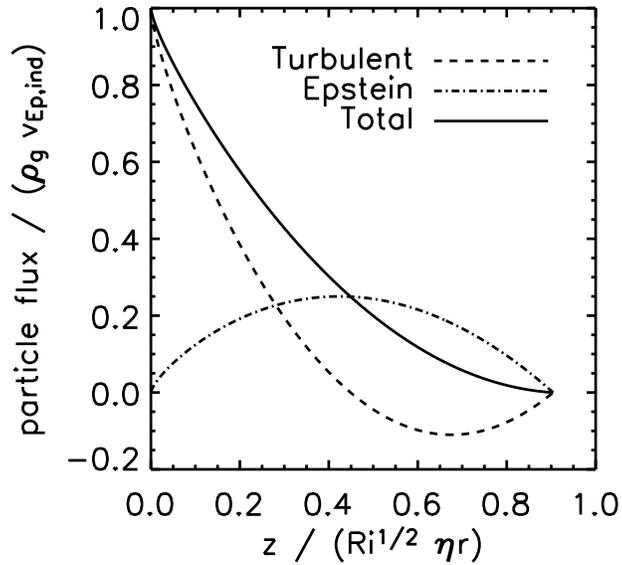, width=9.2cm}
}
\caption{Same as Figure \ref{fluxH1}, except the surface density of solids has
been increased to near
the saturation limit: $\Sigp = 0.99 \Sigpc$.}\label{H1crit}
\end{figure}

\subsubsection{A Self-Consistent Description of Turbulent
Stresses}\label{newcalc}
Proceeding from equation (\ref{LL}), we express the turbulent stress,
\be
\label{Pzp}
P_{z\phi} = \rho \nu_z {\p v_\phi \over \p z}\, ,
\ee
in terms of a momentum diffusivity (viscosity), $\nu_z$,
which controls the efficiency of vertical mixing
of azimuthal momentum.
This same diffusivity also characterizes vertical mixing of particles;
the ratio of the diffusivity of momentum to that
of particles is the Schmidt number, ${\rm Sc}$; it is unity
for the small particles considered here for which $\Omega \ts \ll 1$
and that are entrained in all but the smallest turbulent
eddies \citep{cdc93}.
Given the equivalence of these diffusivities, we may
derive $\nu_z$ from vertical profiles
of particle density
as computed by Sekiya (1998). Our procedure is developed as follows.


The equilibrium distribution of particle density, $\rhop(z)$,
reflects a detailed balance between turbulent diffusion
and gravitational settling. The speed of upward particle
diffusion is given by

\be
\label{diff}
w_{\rm diff} = -\nu_z {\p \ln \rhop \over \p z} \,,
\ee
while the downward vertical settling speed is nearly terminal,
\be
w_{\rm sett} = - \Omega^2 z \ts \, .
\ee
Vertical flux balance,
$w_{\rm diff} + w_{\rm sett} = 0\,$,
allows us to calculate the turbulent viscosity in terms of known quantities:
\be \label{nu}
\nu_z = - {\Omega^2 \ts z \over \p \ln \rhop /\p z} \, .
\ee
Note that the local viscosity increases with stopping time and with
height above the midplane.
More vigorous turbulence is required to keep larger particles
aloft and at greater distances above the midplane where the vertical
component of stellar gravity is stronger.
In addition, strong gradients in particle density
survive only in regions of low viscosity.

We use (\ref{vp}) and (\ref{nu})
in (\ref{Pzp}) to calculate the relevant
component of the stress tensor as
\be
\label{pzprescrip}
P_{z\phi} \approx -\eta r z \Omega^3 \ts {\rhog \over \rho} \, \rhop \, .
\ee
Insertion of (\ref{pzprescrip}) into (\ref{LL}) yields
the resulting equilibrium drift speed,
\be\label{vturb}
\vt \approx -{2 \over \rho \Omega}{\p P_{z\phi} \over \p z} \approx
v_{\rm Ep,ind}{\rhog \over \rho}
{\p \over \p z}\l({\rhop \over \rho}z\r).
\ee
Since the gas scale height, $H\g$, is larger than the particle
scale height ($H\sp/H\g \sim \sqrt{\eta} \ll 1$),
we have ignored vertical gradients in gas properties
in writing equation
(\ref{vturb}).  Note that $\vt$ differs from the Epstein drag rate only
via the shapes of the gas and particle density profiles.
Since $\rhop z/\rho$ is antisymmetric about the midplane
and decreases near the top of
the particle layer (as $\rhop \rightarrow 0$),
equation (\ref{vturb})
implies radial inflow near the midplane and radial
outflow at greater heights.

Note further that with the
exception of the last step in equation (\ref{vturb}), our treatment
in this section does not depend on the validity of the Epstein drag law;
the only requirement is that the Schmidt number be unity, or equivalently
that particles be small enough that $\Omega \ts \ll 1$.
Strict adoption of the results in
Cuzzi et al.~(1993)---see, in particular, their
Figure 2---tells us that the Schmidt number for millimeter-sized particles
in a Hayashi nebula
varies from ${\rm Sc} = 1$ inside $r = 1$ AU to ${\rm Sc} = 10$
at $r = 100$ AU. For particle sizes $a \gtrsim 1$ cm, the Schmidt
number varies more strongly with $r$.
Unfortunately, while it is true asymptotically that
$\Omega \ts \ll 1$ implies ${\rm Sc} = 1$, the precise values for ${\rm Sc}$
cited
in Cuzzi et al.~(1993) depend
on poorly constrained dimensionless numbers; see their sections 2.2
and 2.3. We work, for simplicity, under the assumption
of constant ${\rm Sc} \approx 1$; despite the uncertainties involved,
even this latter requirement could
be relaxed by introducing a spatially variable Schmidt number.

The equilibrium particle mass flux due to turbulent stresses,
\begin{equation}\label{Fturb}
f_{\mathrm{turb}} = \rhop \vt \, ,
\end{equation}
is plotted in Figures \ref{fluxH1} and \ref{H1crit}
for the cases of solar and super-solar metallicities, together
with the previously discussed fluxes due to Epstein drag.
In the case of solar metallicity, the turbulent flux
increases the midplane flux by $\sim 50\%$.
The relative importance of turbulent fluxes
increases with increasing metallicity; for the
metal-rich case displayed in Figure \ref{H1crit}, turbulent
fluxes 
dominate near the midplane.

It appears that turbulent stresses enhance the mass accretion rates due to
Epstein drag.
We proceed to execute global simulations
of radial drift to investigate the possibility of particle pile-ups
in the presence of both Epstein drag and turbulent stresses.

\section{Global Evolution}\label{global}
We solve the mass continuity equation in the radial direction.
Integrated over height, the continuity equation reads
\be\label{cont}
{\p \Sigp \over \p t} = {1 \over r}{\p \over \p r} \l(r F\r) \, ,
\ee
where the total particle flux equals
\be\label{Fint}
F = F_{\rm Ep} + F\t =  \int_{-H\sp}^{H\sp} (f_{\rm Ep} + f\t) \, dz \, .
\ee
As before, we assume an axially symmetric disk
that is everywhere in a state of marginal Kelvin-Helmholtz turbulence.

The integrals of flux over height
can be computed analytically for the Sekiya models
once the midplane density, $\rhop(0)$, is known.  One finds
\bea\label{integ}
{F_{\rm Ep} \over \rhog \eta r v_{\rm Ep,ind}} &=&
\left[ u_0^2 \left( {3 \over 2}\chi - 1 \right) + \chi \psi^2
\right]\mathcal{L}
- \left(1 + {2 \over 3}u_0^2 - 3\chi+ {11 \over 6}\chi^2 \right)\mathcal{R} \\
\label{integ2}
{F\t \over \rhog \eta r v_{\rm Ep,ind}} &=&
 \frac{1}{12}\l\{\l[4u_0^2 + \chi(11\chi-9)\r]
  \mathcal{R} + 3\l[u_0^2(1-3\chi) - 2\psi\chi^2\r]\mathcal{L}\r\}
\eea
where
\bea
u_0 &\equiv& {\rhog \over \rhog + \rhop(0)} + \psi,\\
\chi &\equiv& 1+ \psi \equiv 1 + 4\pi G \rhog /\Omega^2,\\
\mathcal{R} &\equiv& \sqrt{\chi^2 - u_0^2},\\
\mathcal{L} &\equiv& \ln\l(\frac{\chi+\mathcal{R}}{u_0}\r).
\eea
Although cumbersome, these exact expressions permit
greater accuracy and computational efficiency over
direct numerical integration.

Input parameters and boundary values for our first-order partial differential
equation (\ref{cont}) are as follows.
Particles have $\rhos = 3 \gcc$ and $a = 1 \mm$.
The gas disk obeys fixed power laws [equations (\ref{sigg}) and (\ref{T})]
with specific values for indices and normalizations taken from either
model H or model Af (see YS02).
The initial surface density profile for solids is scaled to
that of gas but with an exponential cut-off at large radius,
$\Sigp(r,t=0) = \Sigg \exp{(-x^2)} / 200$,
where $x = r/ (200\AU)$.
The choice of $\Sigp/\Sigg = 5 \times 10^{-3}$ at $x \ll 1$
assumes that only refractory material at solar abundances
has condensed; inclusion of ices would enlarge $\Sigp/\Sigg$.
Our outer boundary condition imposes a ``lid" around the disk's outer edge,
$F(r_{\mathrm{out}} = 350 \AU) = 0$.  We note that an inflow boundary
condition, $\p F/\p r = $ constant at $r = r_{\mathrm{out}}$,
yields nearly identical results because of the
exponential cutoff in $\Sigp$.  No boundary condition is required at the inner
boundary, $r_\mathrm{in} = 0.5 \AU$, where material leaves the grid.

\subsection{Numerical Techniques}
As in all advection problems, we must obey the Courant
condition.  The time step, $\Delta t$, must be smaller than
the time for the flow, at speed $v_r$, to pass from one grid point to the next;
i.e., the Courant number ${\rm Co} \equiv \Delta t v_r/\Delta r < 1$
everywhere, where $\Delta r$ is the grid spacing.
The computational cost is dominated by
regions with small $\Delta r/v$.  Since
$v_r \propto r^d$ roughly, the most efficient (non-adaptive) grid
spacing is $\Delta r \propto r^d$ so that we can fix ${\rm Co} = 0.5$
everywhere.
Such a grid, which is logarithmic only for $d=1$, is generated by the
recursion relation:
\be
r_{j+1} = r_j(1 + \epsilon r_j^{d-1}),
\ee
where the index $j$ labels the gridpoint and $\epsilon$ is a constant
based on the number of radial gridpoints, $N_r$, and the desired
range of $r$-values. This same grid works for all times
because we calculate ${\rm Co}$ using $v_r = v_\mathrm{Ep, ind}$, so that $d$
depends only on time-independent
gas properties; see equation (\ref{depsform}).  Evolutionary increases in
particle density only act to decrease the flow speed, which poses no danger of
violating the Courant condition.

We found (see Figure \ref{compare}) that setting $N_r = 1000$ permitted
fast integration with more than sufficient accuracy.
The finite difference technique for the solution of (\ref{cont}) is
explicit, upwind, forward in time, and first order \citep[see][chap.\
19.1]{nr}.
This simple technique proved convergent, stable, and accurate.

Analytic evaluation of the vertically integrated fluxes
[equations (\ref{integ}) and (\ref{integ2})] requires that we
first obtain $\rhop(z=0)$ given the current
value of $\Sigp$ at every radial gridpoint.  To this end,
we employ a non-linear root finder based on an optimized
M\"uller's method. The root-finding algorithm
requires good initial guesses; it is the rate-limiting step
in our simulation.

We halt the simulation once $\Sigp = \Sigpc$ anywhere. At this point,
$\rhop(0)$ becomes formally infinite, and gravitational instability
ensues. One can imagine continuing the simulation by
removing any excess surface density above $\Sigpc$;
the surface density of agglomerated
planetesimals could then be tracked as a function of position
and time.

\begin{figure}[bt]
\centerline{
\epsfig{figure = 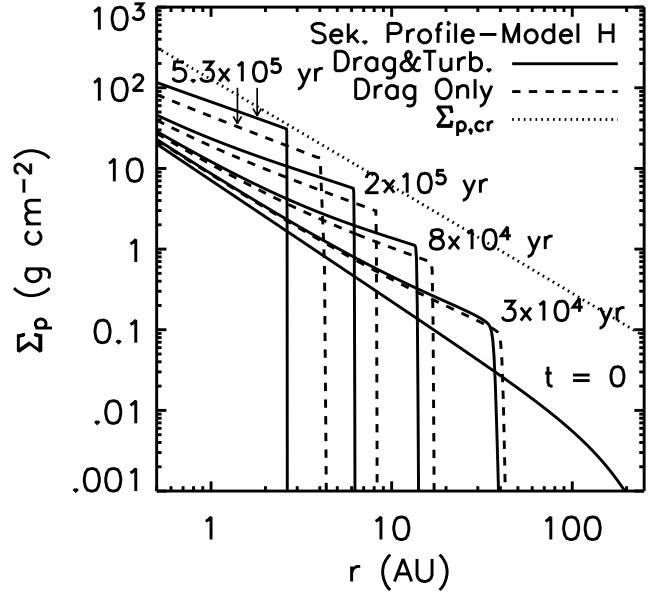, width = 10cm}
}
\caption{Evolution of the surface density, $\Sigp$, of millimeter-sized
particles with time for model H using $10^3$ gridpoints.
Solid lines include both Epstein
drag and turbulent stresses as sources of radial drift; dashed
lines ignore turbulent stresses for comparison.
In both cases, $\Sigp$ grows and its profile flattens with time as the particle
disk contracts radially.
Gravitational instability is first triggered
when $\Sigp$ crosses $\Sigpc$ at the shrinking outer edge, here $r \sim
2.5\,\AU$.}
\label{hev}
\end{figure}

\subsection{Results}\label{res}
We now come to our main result, the evolution of $\Sigp(r,t)$ subject to
Epstein drag and turbulent stresses when the vertical
density profiles, $\rhop(z)$, are supplied by models of \citet{sek98}.
These results improve on the analytic
models of YS02 who considered only Epstein drag and ignored both
the inertial slow-down effect and turbulent stresses.

Figure \ref{hev} showcases the evolution of particle surface density
in model H under two cases: one in which only
Epstein drag (corrected for inertial slow-down)
is included, and a second
in which both Epstein drag and turbulent transport
of angular momentum are included.
Pile-ups of surface density occur whether or not
turbulent stresses are included in the calculation.
Turbulent stresses accelerate the inflow 
and promote particle pile-ups.
The simulation including turbulent stresses was halted
after $5.3 \times 10^5$ yrs
when $\Sigp > \Sigpc$ at $r = 2.5$ AU.
To obtain saturation densities at larger stellocentric radii,
one could start with a larger disk or a more metal-rich disk
in which more solids could be provided by condensible ices.

For model H, planetesimal formation is first triggered at the outer boundary
or ``cliff edge" of the shrinking disk of solids.
This behavior is explained as follows.
If the actual particle surface density
flattens from its initial
profile---i.e., if $|d \ln \Sigp / d\ln r| < |p|$---then
$\Sigp$ will first exceed $\Sigpc$ at the outer boundary
of the accreting disk of particles.
Whether the radial profile of
particle surface density flattens depends on how the timescale
for local density amplification, $\Sigp / \dot{\Sigma}_{\rm p}$,
scales with $r$. Equation (\ref{cont}) reveals
that $\Sigp / \dot{\Sigma}_{\rm p} \propto r/v_r \propto r^{1-d}$.
If we ignore
corrections to $v_r$ from the inertial
slow-down effect and from turbulent stresses,
then $d = d_{\rm Ep} = p = 3/2$ for model H.
Hence, $\Sigp / \dot{\Sigma}_{\rm p} \propto r^{-1/2}$,
the radial profile of $\Sigp$ flattens with time,
and planetesimal formation is triggered first at the outer
boundary of the particle disk.

\begin{figure}[bt]
\centerline{
\epsfig{figure = 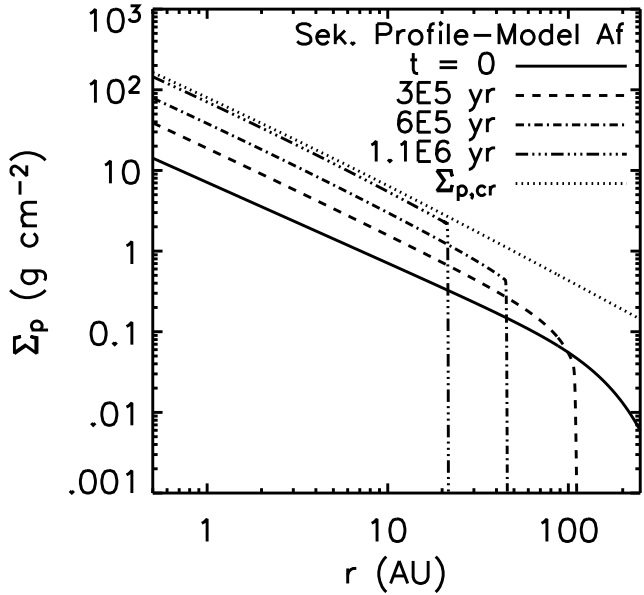, width = 10cm}}
\caption{Similar to Figure \ref{hev}, except for model Af ($p = n = 1$,
$q = 0.63$). Epstein
drag and turbulent stresses are included.  The evolution
proceeds relatively slowly for this model
because the velocity index $d \approx d_{\rm Ep} = 0.87$ ($v_r \propto r^d$)
is smaller than for model H ($d_{\rm{Ep}}=1.5$), so that
drift velocities are smaller for this model than for model H
outside $r \sim 1\AU$. Since
$d_{\rm Ep} < 1$, $\Sigp(r)$ steepens and first exceeds
$\Sigpc$ at the inner boundary of the simulation, $r_\mathrm{in} = 0.5 \AU$.}
\label{af}
\end{figure}

Figure \ref{af} displays the evolution for nebular model Af,
for which initial density profiles are shallower
($p = n = 1$) and temperature profiles are steeper
($q = 0.63$) than for model H. Both Epstein drag and turbulent stresses
are included.  Results that include only Epstein drag
are similar and not shown in Figure \ref{af} so as not to clutter it.
This simulation was halted after $1.2 \times 10^6$ yrs
when $\Sigp > \Sigpc$ at $r = 0.5$ AU, the inner boundary of the
simulation domain. By contrast to model H,
planetesimal formation begins at the inner edge of the solid
disk for model Af; the difference arises because
$d = 0.87 < 1$ for the latter model, so that
$\Sigp / \dot{\Sigma}_{\rm p} \propto r^{+0.13}$
and the radial profile of particle surface density steepens
(see the above discussion for model H).



\begin{figure}[bt]
\centerline{
\epsfig{figure = 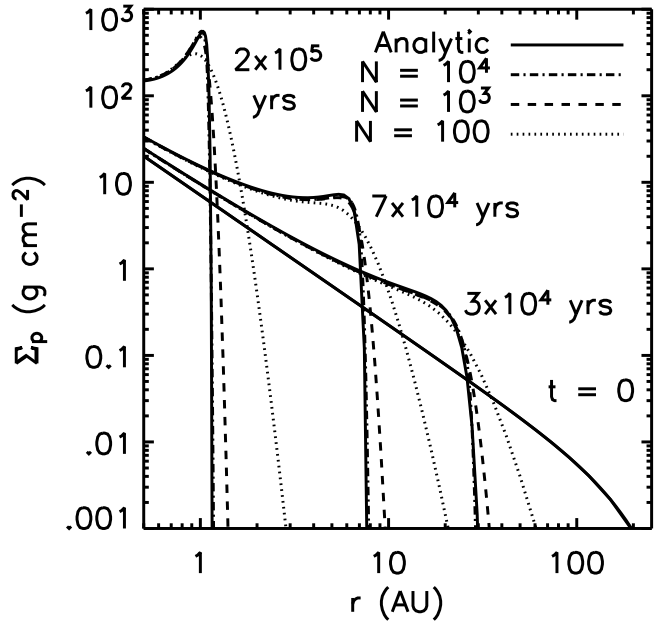,
width = 10cm}
}
\caption{Demonstration that our numerical simulation
converges to the analytic result of YS02,
even when sharp density peaks are present.
Inertial slow-down effects and turbulent stresses are ignored in this
test case; their restoration would smooth away
the density peaks seen here.
The disk is described by model H with chondrule-like particles.  }
\label{compare}
\end{figure}

\subsection{Test Cases}\label{simp}
To gain confidence in our numerical results and to aid in their
interpretation, we ran two test cases that employ
simplified prescriptions for the particle flux.
Since these simple test cases do not depend on the detailed features
of Sekiya's (1998) model, we can also gauge the degree to which our
results are model-independent.
These simple models allow $\Sigp > \Sigpc$ because they are
constructed without regard to the physics of
Kelvin-Helmholtz turbulence, and computationally
they are less intensive.

We first work in the limit of low particle density, $\rhop \ll \rhog$,
and set the particle flux $F = F_{\rm{Ep,ind}} = \Sigp v_{\rm Ep,ind}$.
This expression was used by YS02 to obtain analytic
solutions for the global evolution of particle surface density.
We check our numerical code against these analytic solutions in
Figure \ref{compare} at different grid resolutions.
We chose model H because it gives rise to density cusps
that provide more rigorous tests of numerical resolution
than the smooth density profiles generated by model Af.
The numerical solutions converge to the analytic result
with increasing $N_r$; by $N_r = 10^4$, the analytic and
numerical results are virtually indistinguishable.
(In practice, for the more computationally intensive
models that require root finding, we chose $N_r = 10^3$.)

\begin{figure}[bt]
\centerline{\epsfig{figure = 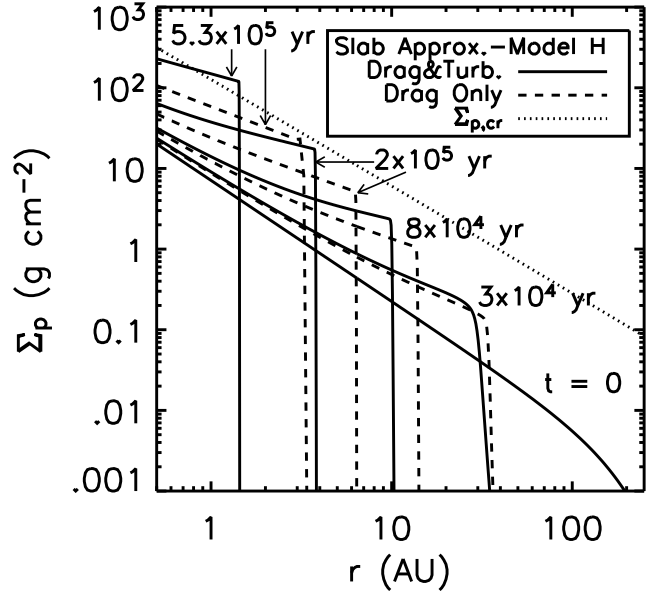, width = 10cm}}
\caption{Similar to Figure \ref{hev}, except $\rhop$ is assumed
to be vertically constant within the particle layer.
This ``slab approximation''
captures the salient features of the full treatment based on
Sekiya's (1998) models, including inertial slow-down,
turbulent speed-up, and the smoothing away
of density peaks seen in Figure \ref{compare}.
This simulation employed $N_r = 10^3$ grid points.}
\label{hslab}
\end{figure}

Our other test case assumes that $\rhop$ is vertically
constant within the particle layer.
This ``slab'' approximation is a useful bridge between the low-density
limit ($F = F_{\mathrm{Ep}}$) and the full treatment based on
the Sekiya profiles. The vertically integrated fluxes
due to Epstein drag and turbulent stresses under the slab approximation
read, respectively,
\bea
\frac{F_{\rm{Ep},\mathrm{slab}}}{2\rhog H\sp v_{\rm Ep,ind}} &=&
\frac{\sigma}{(1+\sigma)^2}~,\\
\frac{F_{\mathrm{turb, slab}}}{2\rhog H\sp v_{\rm Ep,ind}} &=&
\l(\frac{\sigma}{1+\sigma}\r)^2~,
\eea
where $\sigma \equiv \Sigp/(2\rhog H\sp)$.
For comparison, in the limit of low particle density,
$F_{\rm{Ep}}/(2\rhog H\sp v_{\rm Ep}) = \sigma$ increases
without bound with increasing $\Sigp$.
In the slab treatment, accounting for the effect of inertial slow-down
saturates the flux to values below $F_{\rm Ep}$ for
$\sigma \gtrsim 1$.

Figure \ref{hslab} portrays the evolution of particle surface density
in nebular model H under the slab approximation.
The evolution is remarkably similar to that displayed in
Figure \ref{hev}.
In generating Figure \ref{hslab}, we have taken
$H\sp = \sqrt{Ri}\eta r = \eta r / 2$.
In both Figures \ref{hev} and \ref{hslab}, the density
peaks seen in Figure \ref{compare} are smoothed away in similar fashion.
Smoothing occurs because of flux saturation in regions where
$\sigma \gtrsim 1$. Under the slab approximation,
turbulent stresses alter the evolution of particle density
more markedly than under the full treatment;
however, it remains the case qualitatively
that turbulent stresses abet particle pile-ups.

Note that the particle pile-up occurs relatively quickly
in Figure \ref{compare}, while the evolutions shown in
both Figures \ref{hev} and \ref{hslab} evince similar degrees
of inertial slow-down. The slow-down is most pronounced
under the full treatment using the Sekiya profiles.
These differences arise from the fact
that the integral in (\ref{Fint}), when performed at fixed $\Sigp$,
diminishes either when $H\sp$ decreases or when $\rhop(z)$ becomes
more inhomogeneous.

\begin{figure}[bt]
\centerline{\epsfig{figure = 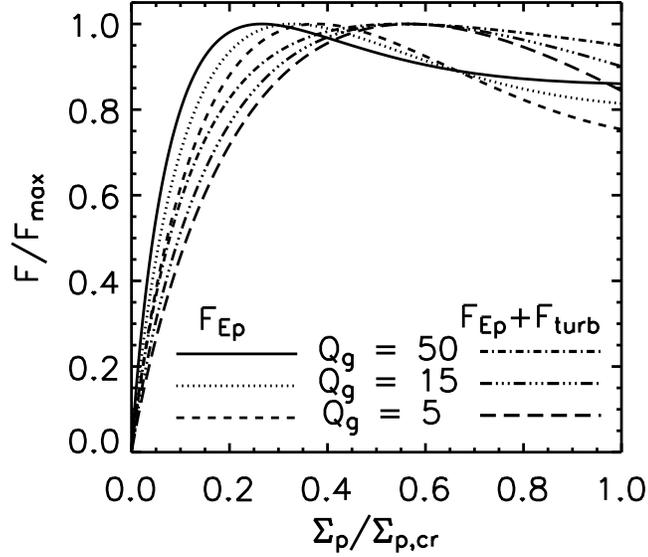, width = 10cm}}
\caption{Vertically integrated, radial particle fluxes
versus $\Sigp$ based on Sekiya's (1998)
profiles.  Results are shown for gas drag with and without
turbulent stresses for several values of Toomre's Q parameter
for the gas, $Q\g$.
Instabilities can occur where $\p F/\p\Sigp < 0$.}
\label{fvssig}
\end{figure}

One qualitative difference arises between using the slab approximation
and the Sekiya models.  In the former case, the radial particle flux
monotonically increases with $\Sigp$ and asymptotes to a constant
value. In the latter case, the flux attains a global maximum at
$\Sigp \approx 0.6 \Sigpc$,
if turbulent stresses and Epstein drag are included.  The maximum occurs
at lower $\Sigp$ if only Epstein drift is considered.
This turnover of the flux is shown in Figure \ref{fvssig}.
At $\Sigp > 0.6 \Sigpc$, the flux decreases with
increasing $\Sigp$. The magnitude of this decrease depends on the strength
of self-gravity of the gas, as measured by the parameter
$Q\g = 4/(\sqrt{2\pi}\psi)$. Smaller values of $Q\g$ indicate stronger
self-gravity and cause a larger drop in the flux. The turn-over occurs
because as $\Sigp$ approaches $\Sigpc$, $H\sp$ decreases and engenders
greater inertial slow-down.
In addition, as $\Sigp$ increases, larger amounts
of solids are swallowed into the cusp in density near the midplane,
where the inertial slow-down effect is greatest.
Both of these behaviors of $\rhop(z)$ with
$\Sigp$ can be seen in Figure 1 of YS02.

The turn-over in $F$ versus $\Sigp$ can spur instabilities.
Imagine that $\p F/\p\Sigp < 0$ and that
a region of high $\Sigp$ exists downstream
of a region of low $\Sigp$.
The low-density region transmits a comparatively large flux to
the high-density region, while the high-density
region parts with a comparatively small flux.
The high-density region therefore amplifies
its density, at the expense of regions yet further downstream.
In this way, the instability propagates downstream.
Indeed, we believe that this instability manifested itself in our simulations
in the form of a ``saw-tooth'' where the density
fluctuated from one radial grid point to the next.
The saw-tooth typically appeared when $\Sigp/\Sigpc \gtrsim 0.8$---$0.9$,
in the regime where $\p F/\p\Sigp < 0$.  Attempts to resolve the
instability at higher grid resolutions were unsuccessful; more study
is needed.\footnote{One could perform a local stability analysis analogous
to that carried out by GP00. This would require restoring the time derivatives
to our momentum equation (\ref{L}).}
While the shape of the $F$-$\Sigp$ relation that we have
computed raises the possibility that the
effective saturation criterion should instead
read $\Sigp > 0.6 \Sigpc$, such a refinement may not be important
in practice. If aerodynamic drift (and other mechanisms)
can raise $\Sigp$ to within a factor of 2 of $\Sigpc$,
they are likely to raise it all the way.


\section{On the Validity of Plate Drag}\label{PD}
In \S\ref{newcalc} and \S\ref{global} we treated turbulent stresses
without recourse to the plate drag approximation.
Since the plate drag formulae are commonly employed in the literature,
we explain here why we believe that
their use is problematic at best and inappropriate at worst.

We have already mentioned in \S\ref{sss:PD} the large range of values
that the critical parameter, ${\rm Re}^*$, has historically taken
and the difficulty in determining its true form.
More fundamentally, the plate drag approximation begs the question of
whether we can
think of a slurry of particles as a monolithic entity.
GP00 propose that such collective behavior is possible if wakes
around particles overlap, analogous to the practice of ``drafting''
in bicycle racing. In the present context of the Epstein drag
regime, these wakes are disturbances having
tiny length scales in the free molecular flow.
If one assumes that each solid
particle leaves a ``footprint'' on the surrounding
fluid flow that extends a distance of order the gas mean free
path, $\lambda$, then one can show that for $\rhop \sim \rhog$
the probability that another particle lies
within this sphere of influence is $\sim$$10^{-8}$.

Despite the failure of free molecular wakes to
overlap, collective effects among solids are not precluded.
Indeed, the ``inertial slow-down'' correction to the particle
drift velocity [equation (\ref{ve})]
tells us that particles are well aware
of each other if their collective density, $\rhop$, approaches $\rhog$.
This communication arises when one accounts for the back-reaction
of particles dragging on gas. We find that this effect describes
most of the collective behavior of solids, even when
turbulent stresses, which are also collective in nature, are included.

Thus, plate drag overestimates turbulent stresses.
Our final criticism concerns the assumption inherent in the
plate drag approximation that the particle-dominated
midplane behaves as a geometrically thin Ekman layer.
We proceed to show that the turbulent diffusivity
in such an Ekman layer is so large that the particle layer cannot be as
thin as assumed.

Ekman boundary layer theory applies to a
medium having viscosity $\nu_{\rm{Ek}}$ and rotating at frequency $\Omega$,
and assumes that other physics (e.g., buoyancy) is negligible.
The height of the Ekman layer is
\be
\label{HEk}
\HEk \approx \sqrt{\nu_{{\rm Ek}}/\Omega} \, .
\ee

\noindent The viscosity, $\nu_{\rm{Ek}}$,
which cannot be directly measured in protoplanetary
disks, is defined in terms of
$\HEk$ and the critical Reynolds number,
\be
\label{heknu}
{\rm Re}^* \equiv \frac{\HEk \eta v_K}{\nu_{\rm{Ek}}} \, ,
\ee

\noindent where the characteristic velocity difference across
the boundary layer is $\Delta v_{\phi} \approx \eta v_K$.
{}From equations (\ref{HEk}) and (\ref{heknu}), we derive
the turbulent diffusivity of the layer,

\be
\label{nuhek}
\nu_{\rm{Ek}} \sim \frac{\eta^2 r v_K}{({\rm Re}^*)^2} \, ,
\ee

\noindent where all quantities, including the critical parameter ${\rm Re}^*$,
are assumed to be known. The corresponding thickness of the Ekman
layer is

\be
\label{hekkle}
H_{\rm Ek} \sim \frac{\eta r}{{\rm Re}^*} \, .
\ee

\noindent Equations (\ref{HEk})--(\ref{hekkle})
are standard in the literature (e.g., \citealp{gw73,dob99};~GP00).

We compare $H_{\rm Ek}$ to the thickness of the particle
layer that is stirred by turbulence
characterized by $\nu_{\rm{Ek}}$. Equation (\ref{nu}) with
$\nu_z = \nu_{\rm Ek}$ yields, to order-of-magnitude, this
thickness\footnote{The assumption from \S\,\ref{newcalc} that $Sc \approx 1$,
true for $\Omega \ts \ll 1$, is still in effect.}:

\be
\label{heckle}
H_{\rm Ek,stir} \sim \sqrt{\frac{\nu_{\rm{Ek}}}{\Omega^2 \ts}} \sim
\frac{H_{\rm Ek}}{\sqrt{\Omega \ts}} \, .
\ee

\noindent Whenever $\Omega \ts \ll 1$,
$H_{\rm Ek,stir}$ is inconsistently larger
than $H_{\rm Ek}$. GP00 appreciate this point (see their equation [14])
and conclude that the plate drag formula yields self-consistent
results for particles large enough (i.e.,
$a \sim 10 \cm$ at $r \sim 1 \AU$) that $\Omega \ts \sim 1$.  If,
however, $\Omega \ts \ll 1$,
then a layer as thin as $H_{\rm Ek}$ will be stirred thicker until
vertical shear becomes too weak to stir it further.
This is the marginally Kelvin-Helmholtz turbulent state
that we have employed throughout this paper.

It should not be too surprising that the turbulent boundary layer
as pictured within the plate drag approximation cannot, in general,
be accurately
described as an Ekman layer. The large Rossby number of the layer,
\be
{\rm Ro} \equiv \frac{\Delta v_\phi}{\Omega \HEk} \approx {\rm Re}^* \gg 1 \, ,
\ee
implies that inertia is more significant than rotation.
Vertical shear and buoyancy are important ingredients
missing from the traditional Ekman layer description.
The Kelvin-Helmholtz turbulent layer that we have
considered has ${\rm Ro} \sim 1$. Recent work by \citet{is03} suggests
that centrifugal and Coriolis forces
weaken vertical shear instabilities.  This finding only
helps to further pave the way for planetesimal
formation by gravitational instability.
See Appendix \ref{modi} for further discussion of rotational
effects on particle drift rates.



Finally, we comment that the Ekman layers in the Earth's atmosphere
are not subject to the crisis described here.
In the Earth's case, atmospheric shear is generated by thermal forcing,
which is unaffected by boundary layer turbulence.  By contrast, vertical
shear in protoplanetary disks is generated by the inertia of particles
within the boundary layer.

In conclusion, in the small particle
limit where $\Omega \ts \ll 1$, the plate drag formula seems
to apply only if particles were effectively glued onto a fixed plate!

\section{Summary}\label{conc}
We have investigated radial drift rates of solid particles within
protoplanetary disks that derive their turbulence from
vertical shear near their strongly stratified midplanes.
Previous models by YS02 calculate drift rates according
to mean flow differences between the sub-Keplerian gas
and the more nearly Keplerian particles. We improve
on their work by incorporating the effects
of collective particle inertia [equation (\ref{ve})]
and the contribution to particle accretion from
turbulent stresses [equation (\ref{vturb})].
These turbulent stresses are calculated in a physically
self-consistent manner; turbulent particle diffusivities
are derived from the detailed balance of vertically upward
and downward particle fluxes within
states of marginal Kelvin-Helmholtz turbulence [as computed
by Sekiya (1998)]. Our results reinforce the conclusions
of YS02 that particle drifts lead to particle pile-ups.
These pile-ups become sites of planetesimal formation
by gravitational instability
once the surface density of solids exceeds the saturation limit.
We have explained why turbulent stresses cannot be modeled
using the traditional plate drag approximation in the (most
interesting) limit of small particle size; our alternative
approach leads us to conclude that turbulent stresses
hasten particle pile-ups.

Inertial slow-down---the diminishing ability of gas to frictionally
sap the angular momentum of particles as the particle mass
density increases above the gas mass density---resulted
in longer accretion timescales and a more even spreading of solids
over larger regions of space, as compared to the results of YS02.
Inclusion of this effect delays the particle pile-up, but only mildly.


How the story of planetesimal formation unfolds in actual circumstellar
disks will depend on a variety of factors in addition
to aerodynamic drift of solids. Photoevaporation by stellar
ultraviolet photons and stellar winds will strip gas from
the disk surface and leave solids near the midplane.
Furthermore, a distribution
of particle sizes and geometries (fluffy vs.~compact)
will smooth particle surface density profiles.

Is planetesimal formation
triggered first near the outer or inner edges of accreting particle
disks? The answer bears directly on the architecture
of planetary systems, on whether the mass in a planetary
system is weighted toward small stellocentric distances (as
they seem to be in extrasolar systems possessing Jovian-mass planets
inside $r \sim 1\AU$) or larger stellocentric distances
(as they are in our solar system). We have quantified the question
of inside-out versus outside-in planet formation
in terms of the index, $d$, which describes the inward
radial drift speed of particles ($v_r \propto r^d$).
If $d > 1$, then particle pile-ups occur first at
the outer edge of the particle disk.
For our power-law disks, $d \simeq d_{\rm{Ep}} = p - q + 1/2$,
where $p$ and $q$ describe the variation of gas surface density
and temperature with disk radius, respectively.
Our knowledge of these quantities is
informed by observations of circumstellar disks
at mid-infrared to millimeter wavelengths.
To date, $q$-values near 0.5 seem difficult
to avoid \citep[see, e.g.,][]{chi01,dch01},
while estimates for $p$-values via dust continuum
observations are traditionally hampered
by lack of spatial resolution and ignorance concerning
the dust opacity and the dust-to-gas ratio.
Improvements in spatial resolution are promised
by future detectors such as the Stratospheric
Observatory for Infrared Astronomy (SOFIA) and
the Atacama Large Millimeter
Array (ALMA), while direct measurements of molecular
hydrogen gas are possible
via its rotational transitions at wavelengths
of 17 and 28 $\mu {\rm m}$ \citep[e.g.,][]{thi99} and its ro-vibrational lines
at 2.1 $\mu \rm{m}$ \citep[e.g.,][]{bwk03}.

As noted by YS02, observations of our solar system's Kuiper Belt
point with ever increasing conviction to the existence of
a hard outer edge to the classical Kuiper Belt at $r \approx 48\AU$
\citep[e.g.,][and references therein]{abm02}.
Were it not for the coincidence of the edge with
the location of the 2:1 mean-motion resonance with Neptune,
we might simply ascribe this edge to a particle pile-up
and the necessity of exceeding a critical threshold
metallicity for planetesimal formation.
We therefore offer a slightly more complicated scenario
to explain these observations.
A hard, primordial edge to the solar system formed
at $r < 48 \AU$ as a consequence of the processes described in YS02
and this work. As Neptune was scattered (stochastically)
outward by planetesimal encounters \citep{fi84,hm99},
its exterior resonances captured bodies and carried
them to greater heliocentric distances \citep{mal95, cj02}.
In this way, the 2:1 (strongest, outermost) Neptunian resonance combed
the primordial edge outward to its present location.

A.\ N.\ Y.\  acknowledges support from a National Science Foundation
Graduate Research Fellowship.
E.I.C. acknowledges support from Hubble Space Telescope Theory Grant
HST-AR-09514.01-A and National Science Foundation Planetary Astronomy
Grant AST-0205892. We thank the referee, Jeremy Goodman, for
a thoughtful report that helped to improve the logic and presentation of this
paper.

\appendix
\section{Modifications to Turbulent Drift Rates by Rotation} \label{modi}
We have modeled Kelvin-Helmholtz turbulence
as occurring in a vertical, Cartesian shear flow, without regard
to the Coriolis force. This force
deflects turbulent eddies in the $z$-$\phi$ plane
into the $r$-$\phi$ plane and thereby
introduces a non-zero $P_{r\phi}$ stress into
equation (\ref{L}) for particle drift rates.
Here we justify the neglect of this term,
$(1/r^2) \partial (r^2 P_{r\phi})/\partial r$,
as compared to $\partial P_{z\phi} / \partial z$.
Since the Rossby number of the turbulent layer
is of order unity,

\begin{equation}
{\rm Ro} = \frac{\Delta v_\phi}{\Omega H\sp} \sim \frac{\eta v_K}{\Omega \eta
r} \sim 1 \, ,
\end{equation}

\noindent the magnitude of $P_{r\phi}$
is the same as that of $P_{z\phi}$.
An equivalent way of seeing this
is to write down an expression for $P_{r\phi}$
analogous to equation (\ref{Pzp}),

\begin{equation}
P_{r\phi} = \rho \nu_r \frac{\partial v_{\phi}}{\partial r} \sim \rho \nu_r
\frac{\Omega r}{r} \sim \rho \nu_r \Omega \, ,
\end{equation}

\noindent which is of the same order as

\begin{equation}
P_{z\phi} = \rho \nu_z \frac{\partial v_{\phi}}{\partial z} \sim \rho \nu_z
\frac{\eta v_K}{\eta r} \sim \rho \nu_z \Omega \, ,
\end{equation}

\noindent since $\nu_r \sim \nu_z$. Then
it is clear that
$(1/r^2) \partial (r^2 P_{r\phi}) / \partial r \sim P_{r\phi} / r$
is smaller than $\partial P_{z\phi} / \partial z \sim P_{z\phi} / H\sp$
by $\sim$$H\sp / r \sim \eta \ll 1$.

\section{Required Levels of Disk Passivity}\label{app}

This appendix quantifies the extent to which circumstellar disks
must be passive for the processes described in this paper to occur.
Here we are considering ``anomalous''
sources of turbulence and angular momentum
transport not arising from vertical shear in stratified particle layers
(e.g., the magneto-rotational instability).
We first ask under what conditions radial accretion velocities arising from
an anomalous viscosity exceed the particle accretion velocities
derived in this paper. For the latter we employ
$v_{\rm{Ep}}$ (since $\vt < v_{\rm{Ep}}$),
and for the former we employ the usual $\alpha$-prescription,

\be
\label{valpha}
v_{\alpha} \sim \alpha_r \eta v_K \, ,
\ee
where $\alpha_r$ parameterizes the strength of radial transport
of angular momentum. Then one necessary criterion for disk
passivity reads

\be \label{ignorealpha}
\alpha_r \ll {\rhog \over \rho}\Omega \ts \, .
\ee
For chondrule-like particles at $r \sim 1 \AU$ in the
MMSN, this criterion is satisfied for
disks with $\alpha_r \ll 10^{-4}$. It could be satisfied in more
active disks for larger particles or at larger disk radii.

A second criterion demands that the disk be sufficiently
quiescent to permit vertical settling of solids.
{}From equation (\ref{nu}), we estimate
the diffusivity generated by Richardson turbulence as
$\nu_z \sim (\eta v_K)^2 \ts$, where we have used $H\sp \sim \eta r$.
We express this diffusivity in terms of an
$\alpha$-viscosity as $\alpha_z \equiv \nu_z\Omega/c_g^2$,
where the subscript $z$ reminds us that turbulence
need not be isotropic. Then this second criterion requires
that any additional source of turbulence apart from
vertical shear be characterized by

\be
\alpha_z \lesssim \Omega\ts \, .
\ee
For our power-law disks containing millimeter-sized particles,
this criterion reads
$\alpha_z \lesssim 10^{-7}\,(r/\rm{AU})^{1+p-q}$.
For model H, $1+p-q=2$.
The most stringent requirements on disk passivity lie
at small stellocentric distances.

We note that $\alpha_z < 10^{-7}$ is still well satisfied
by molecular viscosity, for which
$\alpha_r \sim \alpha_z \sim \lambda/H\g \sim 10^{-12}$
at $r \sim 1\AU$.

Does the anomalous viscosity ever shut off?  Observations of
T Tauri stars (and young brown dwarfs) indicate that infrared excesses
indicating the presence of disks almost always correlate with H$\alpha$
emission and other accretion signatures \citep{kh95, liu03}.  However,
this oft-cited correlation has a considerable amount of scatter.
Figure 4 of \citealp{kh95} clearly shows that a sizable fraction
of systems with infrared excesses have small H$\alpha$ equivalent widths.
These may represent disks that are quiescent enough to form planetesimals.
We note that accretion signatures at optical-to-ultraviolet
wavelengths pertain to activity
in the immediate vicinity of the star; they may implicate
accretion within $\sim$30 stellar radii of the stellar surface,
but they are not diagnostic of conditions at much greater stellocentric
distances. Furthermore, evidence for significant
vertical settling of dust in disk
photospheres and the implied absence of vertical
stirring in the uppermost scale heights is presented by Chiang et al.~(2001),
based on modeling mid--to-far infrared fluxes
of Herbig Ae systems.

In principle, it is possible for disks to exhibit
$\alpha_r \gg \alpha_z$ and to thereby reconcile
modest levels of accretion with vertical settling of solids
and particle pile-ups. For weakly viscous flows having small Rossby number,
the Taylor-Proudman theorem constrains rotating
flow to be two dimensional, parallel to the midplane,
with no variation of velocity with height $z$
\citep[p. 43]{ped87}.
Indeed, in both the Earth's atmosphere
and the Earth's oceans, the ratios of horizontal to vertical
viscosities are roughly estimated to be
$\sim$$10^4$ \citep[p. 185]{ped87}.
By analogy to our problem, we could simultaneously demand
$\alpha_z \sim 10^{-8}$ so as to allow vertical settling and
$\alpha_r \sim 10^{-4}$ so as to yield observable accretion rates
of $\dot{M} \sim 10^{-9}$--$10^{-8} M_\odot/$yr
while not violating (\ref{ignorealpha}) throughout most of the disk.
Thus, mildly accreting class II protostellar systems
could well be sites of planetesimal formation via gravitational instability.


\bibliography{refs}
\end{document}